\documentclass[12pt,notitlepage]{article}\usepackage[]{graphicx}\usepackage[]{color}
\makeatletter
\def\maxwidth{ %
  \ifdim\Gin@nat@width>\linewidth
    \linewidth
  \else
    \Gin@nat@width
  \fi
}
\makeatother

\definecolor{fgcolor}{rgb}{0.345, 0.345, 0.345}
\newcommand{\hlnum}[1]{\textcolor[rgb]{0.686,0.059,0.569}{#1}}%
\newcommand{\hlstr}[1]{\textcolor[rgb]{0.192,0.494,0.8}{#1}}%
\newcommand{\hlcom}[1]{\textcolor[rgb]{0.678,0.584,0.686}{\textit{#1}}}%
\newcommand{\hlopt}[1]{\textcolor[rgb]{0,0,0}{#1}}%
\newcommand{\hlstd}[1]{\textcolor[rgb]{0.345,0.345,0.345}{#1}}%
\newcommand{\hlkwb}[1]{\textcolor[rgb]{0.69,0.353,0.396}{#1}}%
\newcommand{\hlkwd}[1]{\textcolor[rgb]{0.737,0.353,0.396}{\textbf{#1}}}%

\usepackage{framed}
\makeatletter
\newenvironment{kframe}{%
 \def\at@end@of@kframe{}%
 \ifinner\ifhmode%
  \def\at@end@of@kframe{\end{minipage}}%
  \begin{minipage}{\columnwidth}%
 \fi\fi%
 \def\FrameCommand##1{\hskip\@totalleftmargin \hskip-\fboxsep
 \colorbox{shadecolor}{##1}\hskip-\fboxsep
     \hskip-\linewidth \hskip-\@totalleftmargin \hskip\columnwidth}%
 \MakeFramed {\advance\hsize-\width
   \@totalleftmargin\z@ \linewidth\hsize
   \@setminipage}}%
 {\par\unskip\endMakeFramed%
 \at@end@of@kframe}
\makeatother

\definecolor{shadecolor}{rgb}{.97, .97, .97}
\definecolor{messagecolor}{rgb}{0, 0, 0}
\definecolor{warningcolor}{rgb}{1, 0, 1}
\definecolor{errorcolor}{rgb}{1, 0, 0}
\newenvironment{knitrout}{}{} 

\usepackage{alltt}
\usepackage{amsmath}
\usepackage{graphicx,psfrag,epsf}
\usepackage{enumerate}
\usepackage{natbib}
\usepackage{textcomp}
\usepackage[hyphens]{url} 
\usepackage{hyperref}

\usepackage{xspace}
\usepackage{enumitem}

\newcommand{\R}{\textsf{R}\xspace}

\newcommand{\cmd}[1]{\textit{#1}}
\newcommand{\pkg}[1]{\textbf{#1}}
\newcommand{\func}[1]{\textit{#1}()}
\IfFileExists{upquote.sty}{\usepackage{upquote}}{}
\begin{document}

\def\spacingset#1{\renewcommand{\baselinestretch}%
{#1}\small\normalsize} \spacingset{1}

\title{Creating optimal conditions for reproducible data analysis in \R with `fertile'\protect\thanks{The authors gratefully acknowledge contributions from Hadley Wickham, Jenny Bryan, Greg Wilson, Edgar Ruiz, and other members of the \pkg{tidyverse} team.}}

\author{Audrey M. Bertin\\
Statistical and Data Sciences\\
Smith College\\
\and
Benjamin S. Baumer\\
Statistical and Data Sciences\\
Smith College\\
}

\maketitle

\abstract{The advancement of scientific knowledge increasingly depends on ensuring that data-driven research is reproducible: that two people with the same data obtain the same results. However, while the necessity of reproducibility is clear, there are significant behavioral and technical challenges that impede its widespread implementation, and no clear consensus on standards of what constitutes reproducibility in published research. We present \pkg{fertile}, an R package that focuses on a series of common mistakes programmers make while conducting data science projects in R, primarily through the RStudio integrated development environment. \pkg{fertile} operates in two modes: proactively (to prevent reproducibility mistakes from happening in the first place), and retroactively (analyzing code that is already written for potential problems). Furthermore, \pkg{fertile} is designed to educate users on why their mistakes are problematic and how to fix them.}

\section{Introduction}\label{sec1}

Data-based research is considered fully reproducible when the requisite code and data files produce identical results when run by another analyst. As research is becoming increasingly data-driven, and because knowledge can be shared worldwide so rapidly, reproducibility is critical to the advancement of scientific knowledge. Academics around the world have recognized this, and publications and discussions addressing reproducibility appear to have increased in the last several years (\cite{eisner-reproducibility}; \cite{sep-scientific-reproducibility}; \cite{bioessays-gosselin}; \cite{engineering-reproducibility}; \cite{plos-biology}). 

Reproducible research has a wide variety of benefits. When researchers provide the code and data used for their work in a well-organized and reproducible format, readers are more easily able to determine the veracity of any findings by following the steps from raw data to conclusions. The creators of reproducible research can also more easily receive more specific feedback (including bug fixes) on their work. Moreover, others interested in the research topic can use the code to apply the methods and ideas used in one project to their own work with minimal effort.

However, while the necessity of reproducibility is clear, there are significant behavioral and technical challenges that impede its widespread implementation and no clear consensus on standards of what constitutes reproducibility in published research (\cite{peng2009bio}). Not only are the components of reproducible research up for discussion (e.g., need the software be open source?), but the corresponding recommendations for ensuring reproducibility also vary (e.g., should raw and processed data files be in separate directories?).
Existing attempts to address reproducibility in data science are often either too generalized---resulting in shallow and vague recommendations that are challenging to implement---or too specific---approaching one aspect of reproducibility well but doing so in a highly technical way that fails to capture the bigger picture and creates challenges for inexperienced users.

In order to address these challenges, we present \pkg{fertile}(\cite{R-fertile}), a low barrier-to-entry package focusing on common reproducibility mistakes programmers make while conducting data science research in R.

\section{Related Work}\label{sec:lit}

\subsection{Literature Review}\label{sec:papers}

Publications on reproducibility can be found in all areas of scientific research. However, as \cite{Goodman341ps12} argue, the language and conceptual framework of research reproducibility varies significantly across the sciences, and there are no clear standards on reproducibility agreed upon by the scientific community as a whole. We consider recommendations from a variety of fields and determine the key aspects of reproducibility faced by scientists in different disciplines. 

\cite{kitzes2017practice} present a collection of case studies on reproducibility practices from across the data-intensive sciences, illustrating a variety of recommendations and techniques for achieving reproducibility. Although their work does not come to a consensus on the exact standards of reproducibility that should be followed, several common trends and principles emerge from their case studies: 1) use clear separation, labeling, and documentation, 2) automate processes when possible, and 3) design the data analysis workflow as a sequence of small steps glued together, with outputs from one step serving as inputs into the next. This is a common suggestion within the computing community, originating as part of the Unix philosophy (\cite{unix}).

\cite{cooper2017guide} focus on data analysis in \R and identify a similar list of important reproducibility components, reinforcing the need for clearly labeled, well-documented, and well-separated files. In addition, they recommend publishing a list of dependencies and using version control. 
\cite{broman} reiterates the need for clear naming and file separation while sharing several additional suggestions: keep the project contained in one directory, use relative paths, and include a \cmd{README}.

The reproducibility recommendations from R OpenSci, a non-profit initiative founded in 2011 to make scientific data retrieval reproducible, share similar principles to those discussed previously. They focus on a need for a well-developed file system, with no extraneous files and clear labeling. They also reiterate the need to note dependencies and use automation when possible, while making clear a suggestion not present in the previously-discussed literature: the need to use seeds, which allow for the saving and restoring of the random number generator state, when running code involving randomness (\cite{r-opensci}).

When considered in combination, these sources provide a well-rounded picture of the components important to research reproducibility. Using this literature as a guideline, we identify several key features of reproducible work. These recommendations are a matter of opinion---due to the lack of agreement on which components of reproducibility are most important, we select those that are mentioned most often, as well as some that are mentioned less but that we view as important. 
\vspace{3mm}

\begin{enumerate} [nolistsep]
\item A well-designed file structure:
  \begin{itemize}[noitemsep]
  \item Separate folders for different file types.
  \item No extraneous files.
  \item Minimal clutter.
  \end{itemize}
\item Good documentation:
  \begin{itemize} [noitemsep]
  \item Files are clearly named, preferably in a way where the order in which they should be run is clear.
  \item A README is present.
  \item Dependencies are noted.
  \end{itemize}
\item Reproducible file paths:
  \begin{itemize}
  \item No absolute paths, or paths leading to locations outside of a project's directory, are used in code---only portable (relative) paths.
  \end{itemize}
\item Randomness is accounted for:
  \begin{itemize}
  \item If randomness is used in code, a seed must also be set.
  \end{itemize}
\item Readable, styled code:
  \begin{itemize}
  \item Code should be written in a coherent style. Code that conforms to a style guide or is written in a consistent dialect is easier to read (\cite{hermans2017programming}). We believe that the \pkg{tidyverse} provides the most accessible dialect of \R.
  \end{itemize}
\end{enumerate}

Much of the available literature focuses on file structure, organization, and naming, and \pkg{fertile}'s features are consistent with this. \cite{marwick2018packaging} provide the framework for file structure that \pkg{fertile} is based on: a structure similar to that of an \R package (\cite{coreteam-extensions}, \cite{hadley-packages}), with an \cmd{R} folder, as well as \cmd{data}, \cmd{data-raw}, \cmd{inst}, and \cmd{vignettes}.

\subsection{R Packages and Other Software}

Much of the work discussed in Section \ref{sec:papers} is highly generalized, written to be applicable to users working with a variety of statistical software programs. Because all statistical software programs operate differently, these recommendations are inherently vague and difficult to implement, particularly to new analysts who are relatively unfamiliar with their software. Focused attempts to address reproducibility in specific certain software programs are more likely to be successful. We focus on \R, due to its open-source nature, accessibility, and popularity as a tool for statistical analysis.

A small body of \R packages focuses on research reproducibility. \pkg{rrtools} (\cite{R-rrtools}) addresses some of the issues discussed in \cite{marwick2018packaging} by creating a basic \R package structure for a data analysis project and implementing a basic \func{testthat::check} functionality. The \pkg{orderly} (\cite{R-orderly}) package also focuses on file structure, requiring the user to declare a desired project structure (typically a step-by-step structure, where outputs from one step are inputs into the next) at the beginning and then creating the files necessary to achieve that structure. \pkg{workflowr}'s (\cite{R-workflowr}) functionality is based around version control and making code easily available online. It works to generate a website containing time-stamped, versioned, and documented results. \pkg{checkers} (\cite{R-checkers}) allows you to create custom checks that examine different aspects of reproducibility. \pkg{packrat} (\cite{R-packrat}) is focused on dependencies, creating a packaged folder containing a project as well as all of its dependencies so that projects dependent on lesser-used packages can be easily shared across computers. \pkg{drake} (\cite{R-drake}) analyzes workflows, skips steps where results are up to date, and provides evidence that results match the underlying code and data. Lastly, the \pkg{reproducible} (\cite{R-reproducible}) package focuses on the concept of caching: saving information so that projects can be run faster each time they are re-completed from the start.

Many of these packages are narrow, with each effectively addressing a small component of reproducibility: file structure, modularization of code, version control, etc. These packages often  succeed in their area of focus, but at the cost of accessibility to a wider audience. Their functions are often quite complex to use, and many steps must be completed to achieve the required reproducibility goal. This cumbersome nature means that most reproducibility packages currently available are not easily accessible to users near the beginning of their \R journey, nor particularly useful to those looking for quick and easy reproducibility checks.
A more effective way of realizing widespread reproducibility is to make the process for doing so simple enough that it takes little to no conscious effort to implement. You want users to "fall into a hole"\footnote{We paraphrase Hadley Wickham.} of good practice. 

\emph{Continuous integration} tools provide more general approaches to automated checking, which can enhance reproducibility with minimal code. For example, \pkg{wercker}---a command line tool that leverages Docker---enables users to test whether their projects will successfully compile when run on a variety of operating systems without access to the user's local hard drive (\cite{wercker}). \pkg{GitHub Actions} is integrated into GitHub and can be configured to do similar checks on projects hosted in repositories. \href{https://travis-ci.com}{\pkg{Travis CI}} and \href{https://circleci.com/product}{\pkg{Circle CI}} are popular continuous integration tools that can also be used to check \R code. 

However, while these tools can be useful, they are generalized so as to be useful to the widest audience. As a result, their checks are not designed to be \R-specific, which makes them sub-optimal for users looking to address reproducibility issues involving features specific to the \R programming language, such as package installation and seed setting.

\subsection{Our contribution}

\pkg{fertile} attempts to address these gaps in existing software by providing a simple, easy-to-learn reproducibility package that, rather than focusing intensely on a specific area, provides some information about a wide variety of aspects influencing reproducibility. \pkg{fertile} is flexible, offering benefits to users at any stage in the data analysis workflow, and provides \R-specific features, which address certain aspects of reproducibility that can be missed by external project development software.

\pkg{fertile} is designed to be used on data analyses organized as \R Projects (i.e. directories containing an \cmd{.Rproj} file). Once an \R Project is created, \pkg{fertile} provides benefits throughout the data analysis process, both during development as well as after the fact. \pkg{fertile} achieves this by operating in two modes: proactively (to prevent reproducibility mistakes from happening in the first place), and retroactively (analyzing code that has already been written for potential problems).

\section{Methods}\label{sec3}

\subsection{Proactive Use}

Proactively, the package identifies potential mistakes as they are made by the user and outputs an informative message as well as a recommended solution. For example, \pkg{fertile} catches when a user passes a potentially problematic file path---such as an absolute path, or a path that points to a location outside of the project directory---to a variety of common input/output functions operating on many different file types.

\begin{knitrout}
\definecolor{shadecolor}{rgb}{0.969, 0.969, 0.969}\color{fgcolor}\begin{kframe}
\begin{alltt}
\hlkwd{library}\hlstd{(fertile)}
\hlkwd{file.exists}\hlstd{(}\hlstr{"~/Desktop/my_data.csv"}\hlstd{)}
\end{alltt}
\begin{verbatim}
## [1] TRUE
\end{verbatim}
\begin{alltt}
\hlkwd{read.csv}\hlstd{(}\hlstr{"~/Desktop/my_data.csv"}\hlstd{)}
\end{alltt}

{\ttfamily\noindent\bfseries\color{errorcolor}{\#\# Error: Detected absolute paths}}\begin{alltt}
\hlkwd{read.csv}\hlstd{(}\hlstr{"../../../Desktop/my_data.csv"}\hlstd{)}
\end{alltt}

{\ttfamily\noindent\bfseries\color{errorcolor}{\#\# Error: Detected paths that lead outside the project directory}}\end{kframe}
\end{knitrout}

\pkg{fertile} is even more aggressive with functions (like \func{setwd}) that are almost certain to break reproducibility, causing them to throw errors that prevent their execution and providing recommendations for better alternatives.

\begin{knitrout}
\definecolor{shadecolor}{rgb}{0.969, 0.969, 0.969}\color{fgcolor}\begin{kframe}
\begin{alltt}
\hlkwd{setwd}\hlstd{(}\hlstr{"~/Desktop"}\hlstd{)}
\end{alltt}

{\ttfamily\noindent\bfseries\color{errorcolor}{\#\# Error: setwd() is likely to break reproducibility. Use here::here() instead.}}\end{kframe}
\end{knitrout}

These proactive warning features are activated immediately after attaching the \pkg{fertile} package and require no additional effort by the user.

\subsection{Retroactive Use}

Retroactively, \pkg{fertile} analyzes potential obstacles to reproducibility in an RStudio Project (i.e., a directory that contains an \cmd{.Rproj} file). The package considers several different aspects of the project which may influence reproducibility, including the directory structure, file paths, and whether randomness is used thoughtfully.
The end products of these analyses are reproducibility reports summarizing a project's adherence to reproducibility standards and recommending remedies for where the project falls short. For example, \pkg{fertile} might identify the use of randomness in code and recommend setting a seed if one is not present.

Users can access the majority of \pkg{fertile}'s retroactive features through two primary functions, \func{proj\_check} and \func{proj\_analyze}. 

The \func{proj\_check} function runs fifteen different reproducibility tests, noting which ones passed, which ones failed, the reason for failure, a recommended solution, and a guide to where to look for help. These tests include: looking for a clear build chain, checking to make sure the root level of the project is clear of clutter, confirming that there are no files present that are not being directly used by or created by the code, and looking for uses of randomness that do not have a call to \func{set.seed} present. A full list is provided below:

\begin{knitrout}
\definecolor{shadecolor}{rgb}{0.969, 0.969, 0.969}\color{fgcolor}\begin{kframe}
\begin{alltt}
\hlkwd{list_checks}\hlstd{()}
\end{alltt}

{\ttfamily\noindent\itshape\color{messagecolor}{\#\# -- The available checks in `fertile` are as follows: ----------------------}}\begin{verbatim}
##  [1] "has_tidy_media"          "has_tidy_images"        
##  [3] "has_tidy_code"           "has_tidy_raw_data"      
##  [5] "has_tidy_data"           "has_tidy_scripts"       
##  [7] "has_readme"              "has_no_lint"            
##  [9] "has_proj_root"           "has_no_nested_proj_root"
## [11] "has_only_used_files"     "has_clear_build_chain"  
## [13] "has_no_absolute_paths"   "has_only_portable_paths"
## [15] "has_no_randomness"
\end{verbatim}
\end{kframe}
\end{knitrout}

Subsets of the fifteen tests can be invoked using the \pkg{tidyselect} helper functions (\cite{R-tidyselect}) in combination with the more limited \func{proj\_check\_some} function.

\begin{knitrout}
\definecolor{shadecolor}{rgb}{0.969, 0.969, 0.969}\color{fgcolor}\begin{kframe}
\begin{alltt}
\hlstd{proj_dir} \hlkwb{<-} \hlstr{"project_miceps"}
\end{alltt}
\end{kframe}
\end{knitrout}

\begin{knitrout}
\definecolor{shadecolor}{rgb}{0.969, 0.969, 0.969}\color{fgcolor}\begin{kframe}
\begin{alltt}
\hlkwd{proj_check_some}\hlstd{(proj_dir,} \hlkwd{contains}\hlstd{(}\hlstr{"paths"}\hlstd{))}
\end{alltt}

{\ttfamily\noindent\itshape\color{messagecolor}{\#\# -- Compiling... ------------------------------------- fertile 0.0.0.9027 --}}

{\ttfamily\noindent\itshape\color{messagecolor}{\#\# -- Rendering R scripts... --------------------------- fertile 0.0.0.9027 --}}

{\ttfamily\noindent\itshape\color{messagecolor}{\#\# -- Running reproducibility checks ------------------- fertile 0.0.0.9027 --}}

{\ttfamily\noindent\itshape\color{messagecolor}{\#\# v Checking for no absolute paths}}

{\ttfamily\noindent\itshape\color{messagecolor}{\#\# v Checking for only portable paths}}\end{kframe}
\begin{kframe}

{\ttfamily\noindent\itshape\color{messagecolor}{\#\# -- Summary of fertile checks ------------------------ fertile 0.0.0.9027 --}}\end{kframe}
\begin{kframe}

{\ttfamily\noindent\itshape\color{messagecolor}{\#\# v Reproducibility checks passed: 2}}\end{kframe}
\end{knitrout}

Each test can also be run individually by calling the function matching its check name.

The \func{proj\_analyze} function creates a report documenting the structure of a data analysis project. This report contains information about all packages referenced in code, the files present in the directory and their types, suggestions for moving files to create a more organized structure, and a list of reproducibility-breaking file paths used in code.

\begin{knitrout}
\definecolor{shadecolor}{rgb}{0.969, 0.969, 0.969}\color{fgcolor}\begin{kframe}
\begin{alltt}
\hlkwd{proj_analyze}\hlstd{(proj_dir)}
\end{alltt}

{\ttfamily\noindent\itshape\color{messagecolor}{\#\# -- Analysis of reproducibility for project\_miceps --- fertile 0.0.0.9027 --}}

{\ttfamily\noindent\itshape\color{messagecolor}{\#\# --\ \  Packages referenced in source code ------------- fertile 0.0.0.9027 --}}\begin{verbatim}
## # A tibble: 9 x 3
##   package       N used_in                    
##   <chr>     <int> <chr>                      
## 1 broom         1 project_miceps/analysis.Rmd
## 2 dplyr         1 project_miceps/analysis.Rmd
## 3 ggplot2       1 project_miceps/analysis.Rmd
## 4 purrr         1 project_miceps/analysis.Rmd
## 5 readr         1 project_miceps/analysis.Rmd
## 6 rmarkdown     1 project_miceps/analysis.Rmd
## 7 skimr         1 project_miceps/analysis.Rmd
## 8 stargazer     1 project_miceps/analysis.Rmd
## 9 tidyr         1 project_miceps/analysis.Rmd
\end{verbatim}

{\ttfamily\noindent\itshape\color{messagecolor}{\#\# --\ \  Files present in directory --------------------- fertile 0.0.0.9027 --}}\begin{verbatim}
## # A tibble: 9 x 4
##   file               ext        size mime                                       
##   <fs::path>         <chr> <fs::byt> <chr>                                      
## 1 Estrogen_Receptor~ docx     10.97K application/vnd.openxmlformats-officedocum~
## 2 citrate_v_time.png png     188.29K image/png                                  
## 3 proteins_v_time.p~ png     377.95K image/png                                  
## 4 Blot_data_updated~ csv      14.43K text/csv                                   
## 5 CS_data_redone.csv csv       7.39K text/csv                                   
## 6 mice.csv           csv      14.33K text/csv                                   
## 7 README.md          md           39 text/markdown                              
## 8 miceps.Rproj       Rproj       204 text/rstudio                               
## 9 analysis.Rmd       Rmd       4.94K text/x-markdown
\end{verbatim}

{\ttfamily\noindent\itshape\color{messagecolor}{\#\# --\ \  Suggestions for moving files ------------------- fertile 0.0.0.9027 --}}\begin{verbatim}
## # A tibble: 7 x 3
##   path_rel           dir_rel    cmd                                             
##   <fs::path>         <fs::path> <chr>                                           
## 1 Blot_data_updated~ data-raw   file_move('project_miceps/Blot_data_updated.csv~
## 2 CS_data_redone.csv data-raw   file_move('project_miceps/CS_data_redone.csv', ~
## 3 Estrogen_Receptor~ inst/other file_move('project_miceps/Estrogen_Receptors.do~
## 4 analysis.Rmd       vignettes  file_move('project_miceps/analysis.Rmd', fs::di~
## 5 citrate_v_time.png inst/image file_move('project_miceps/citrate_v_time.png', ~
## 6 mice.csv           data-raw   file_move('project_miceps/mice.csv', fs::dir_cr~
## 7 proteins_v_time.p~ inst/image file_move('project_miceps/proteins_v_time.png',~
\end{verbatim}

{\ttfamily\noindent\itshape\color{messagecolor}{\#\# --\ \  Problematic paths logged ----------------------- fertile 0.0.0.9027 --}}\begin{verbatim}
## NULL
\end{verbatim}
\end{kframe}
\end{knitrout}

\subsection{Logging}

\pkg{fertile} also contains logging functionality, which records commands run in the console that have the potential to affect reproducibility, enabling users to look at their past history at any time. The package focuses mostly on package loading and file opening, noting which function was used, the path or package it referenced, and the timestamp at which that event happened. Users can access the log recording their commands at any time via the \func{log\_report} function:

\begin{knitrout}
\definecolor{shadecolor}{rgb}{0.969, 0.969, 0.969}\color{fgcolor}\begin{kframe}
\begin{alltt}
\hlkwd{log_report}\hlstd{()}
\end{alltt}
\begin{verbatim}
## # A tibble: 3 x 4
##   path         path_abs                             func     timestamp          
##   <chr>        <chr>                                <chr>    <dttm>             
## 1 package:pur~ <NA>                                 base::l~ 2020-08-18 18:46:25
## 2 package:for~ <NA>                                 base::l~ 2020-08-18 18:46:25
## 3 project_mic~ /home/bbaumer/Dropbox/git/fertile-p~ readr::~ 2020-08-18 18:46:25
\end{verbatim}
\end{kframe}
\end{knitrout}

The log, if not managed, can grow very long over time. For users who do not desire such functionality, \func{log\_clear} provides a way to erase the log and start over.

\subsection{How It Works}

Much of the functionality in \pkg{fertile} is achieved by writing \href{https://en.wikipedia.org/wiki/Shim_(computing)}{shims}. \pkg{fertile}'s shimmed functions intercept the user's commands and perform various logging and checking tasks before executing the desired function. Our process is:

\begin{enumerate}[noitemsep]
\item Identify an \R function that is likely to be involved in operations that may break reproducibility. Popular functions associated with only one package (e.g., \func{read\_csv} from \pkg{readr}) are ideal candidates.
\item Create a function in \pkg{fertile} with the same name that takes the same arguments (and always the dots \cmd{...}).
\item Write this new function so that it: a) captures any arguments, b) logs the name of the function called, c) performs any checks on these arguments, and d) calls the original function with the original arguments. Except where warranted, the execution looks the same to the user as if they were calling the original function.
\end{enumerate}
Most shims are quite simple and look something like what is shown below for \func{read\_csv}. 

\begin{knitrout}
\definecolor{shadecolor}{rgb}{0.969, 0.969, 0.969}\color{fgcolor}\begin{kframe}
\begin{alltt}
\hlstd{fertile}\hlopt{::}\hlstd{read_csv}
\end{alltt}
\begin{verbatim}
## function (file, ...) 
## {
##     if (interactive_log_on()) {
##         log_push(file, "readr::read_csv")
##         check_path_safe(file)
##         readr::read_csv(file, ...)
##     }
## }
## <bytecode: 0x55efa4658478>
## <environment: namespace:fertile>
\end{verbatim}
\end{kframe}
\end{knitrout}

\pkg{fertile} shims many common functions, including those that read in a variety of data types, write data, and load packages. This works both proactively and retroactively, as the shimmed functions written in \pkg{fertile} are activated both when the user is coding interactively and when a file containing code is rendered.

In order to ensure that the \pkg{fertile} versions of functions ("shims") always supersede ("mask") their original namesakes when called, \pkg{fertile} uses its own shims of the \func{library} and \func{require} functions to  manipulate the \R \func{search} path so that it is always located in the first position. In the \pkg{fertile} version of \func{library}, we detach \pkg{fertile} from the search path, load the requested package, and then re-attach \pkg{fertile}. This ensures that when a user executes a command, \R will check \pkg{fertile} for a matching function before considering other packages. While it is possible that this shifty behavior could lead to unintended consequences, our goal is to catch a good deal of problems before they become problematic. Users can easily disable \pkg{fertile} by detaching it, or not loading it in the first place.

\subsection{Utility Functions}

\pkg{fertile} also provides several useful utility functions that may assist with the process of data analysis. 

\subsubsection{File Paths}

The \func{check\_path} function analyzes a vector of paths (or a single path) to determine whether there are any absolute paths or paths that lead outside the project directory.

\begin{knitrout}
\definecolor{shadecolor}{rgb}{0.969, 0.969, 0.969}\color{fgcolor}\begin{kframe}
\begin{alltt}
\hlcom{# Path inside the directory}
\hlkwd{check_path}\hlstd{(}\hlstr{"project_miceps"}\hlstd{)}
\end{alltt}
\begin{verbatim}
## # A tibble: 0 x 3
## # ... with 3 variables: path <chr>, problem <chr>, solution <chr>
\end{verbatim}
\begin{alltt}
\hlcom{# Absolute path (current working directory)}
\hlkwd{check_path}\hlstd{(}\hlkwd{getwd}\hlstd{())}
\end{alltt}

{\ttfamily\noindent\bfseries\color{errorcolor}{\#\# Error: Detected absolute paths}}\begin{alltt}
\hlcom{# Path outside the directory}
\hlkwd{check_path}\hlstd{(}\hlstr{"../fertile.Rmd"}\hlstd{)}
\end{alltt}

{\ttfamily\noindent\bfseries\color{errorcolor}{\#\# Error: Detected paths that lead outside the project directory}}\end{kframe}
\end{knitrout}

\subsubsection{File Types}

There are several functions that can be used to check the type of a file:

\begin{knitrout}
\definecolor{shadecolor}{rgb}{0.969, 0.969, 0.969}\color{fgcolor}\begin{kframe}
\begin{alltt}
\hlkwd{is_data_file}\hlstd{(fs}\hlopt{::}\hlkwd{path}\hlstd{(proj_dir,} \hlstr{"mice.csv"}\hlstd{))}
\end{alltt}
\begin{verbatim}
## [1] TRUE
\end{verbatim}
\begin{alltt}
\hlkwd{is_image_file}\hlstd{(fs}\hlopt{::}\hlkwd{path}\hlstd{(proj_dir,} \hlstr{"proteins_v_time.png"}\hlstd{))}
\end{alltt}
\begin{verbatim}
## [1] TRUE
\end{verbatim}
\begin{alltt}
\hlkwd{is_text_file}\hlstd{(fs}\hlopt{::}\hlkwd{path}\hlstd{(proj_dir,} \hlstr{"README.md"}\hlstd{))}
\end{alltt}
\begin{verbatim}
## [1] TRUE
\end{verbatim}
\begin{alltt}
\hlkwd{is_r_file}\hlstd{(fs}\hlopt{::}\hlkwd{path}\hlstd{(proj_dir,} \hlstr{"analysis.Rmd"}\hlstd{))}
\end{alltt}
\begin{verbatim}
## [1] TRUE
\end{verbatim}
\end{kframe}
\end{knitrout}

\subsubsection{Temporary Directories}

The \func{sandbox} function allows the user to make a copy of their project in a temporary directory. This can be useful for ensuring that projects run properly when access to the local file system is removed.

\begin{knitrout}
\definecolor{shadecolor}{rgb}{0.969, 0.969, 0.969}\color{fgcolor}\begin{kframe}
\begin{alltt}
\hlstd{proj_dir}
\end{alltt}
\begin{verbatim}
## [1] "project_miceps"
\end{verbatim}
\begin{alltt}
\hlstd{fs}\hlopt{::}\hlkwd{dir_ls}\hlstd{(proj_dir)} \hlopt{%>%} \hlkwd{head}\hlstd{(}\hlnum{3}\hlstd{)}
\end{alltt}
\begin{verbatim}
## project_miceps/Blot_data_updated.csv   project_miceps/CS_data_redone.csv      
## project_miceps/Estrogen_Receptors.docx
\end{verbatim}
\begin{alltt}
\hlstd{temp_dir} \hlkwb{<-} \hlkwd{sandbox}\hlstd{(proj_dir)}
\hlstd{temp_dir}
\end{alltt}
\begin{verbatim}
## /tmp/Rtmpe7cKju/project_miceps
\end{verbatim}
\begin{alltt}
\hlstd{fs}\hlopt{::}\hlkwd{dir_ls}\hlstd{(temp_dir)} \hlopt{%>%} \hlkwd{head}\hlstd{(}\hlnum{3}\hlstd{)}
\end{alltt}
\begin{verbatim}
## /tmp/Rtmpe7cKju/project_miceps/Blot_data_updated.csv
## /tmp/Rtmpe7cKju/project_miceps/CS_data_redone.csv
## /tmp/Rtmpe7cKju/project_miceps/Estrogen_Receptors.docx
\end{verbatim}
\end{kframe}
\end{knitrout}

\subsubsection{Managing Project Dependencies}

One of the challenges with ensuring that work is reproducible is the issue of dependencies. Many data analysis projects reference a variety of \R packages in their code. When such projects are shared with other users who may not have the required packages downloaded, it can cause errors that prevent the project from running properly. 

The \func{proj\_pkg\_script} function assists with this issue by making it simple and fast to download dependencies. When run on an \R project directory, the function creates a \cmd{.R} script file that contains the code needed to install all of the packages referenced in the project, differentiating between packages located on CRAN and those located on GitHub.

\begin{knitrout}
\definecolor{shadecolor}{rgb}{0.969, 0.969, 0.969}\color{fgcolor}\begin{kframe}
\begin{alltt}
\hlstd{install_script} \hlkwb{<-} \hlkwd{proj_pkg_script}\hlstd{(proj_dir)}
\hlkwd{cat}\hlstd{(}\hlkwd{readChar}\hlstd{(install_script,} \hlnum{1e5}\hlstd{))}
\end{alltt}
\begin{verbatim}
## # Run this script to install the required packages for this R project.
## # Packages hosted on CRAN...
## install.packages(c( 'broom', 'dplyr', 'ggplot2', 'purrr', 'readr', 'rmarkdown', 'skimr', 'stargazer', 'tidyr' ))
## # Packages hosted on GitHub...
\end{verbatim}
\end{kframe}
\end{knitrout}

\subsection{Sample Use Cases}

\pkg{fertile}'s simplicity enables users of any background to take advantage of its features and its big-picture design gives \pkg{fertile} the potential to provide benefits across a variety of disciplines. 

For example, professors could integrate \pkg{fertile} into their data science curricula, giving students an understanding and awareness of reproducibility early in their careers that can positively impact the reproducibility of their future work. It could also be used by experienced analysts working collaboratively who are looking to promote a smoother exchange of feedback and ideas. Journal reviewers may also find the package beneficial, allowing them to gain a fast overview of whether paper submissions meet reproducibility guidelines.

The sample use cases in this section consider \pkg{fertile}'s applicability to some of these scenarios in detail.

\subsubsection{Introductory Data Science Student}

Susan is taking an introductory data science course. This is her first time learning how to code and she has not yet been exposed to ideas of research reproducibility. Her professor has assigned a data analysis project that must be completed in \R Markdown. The project requires her to read in a data file located on her computer and use it to produce a graph.

She reads in the data, makes the graph, and knits her \cmd{.Rmd} file. It compiles successfully, so she submits the assignment. The next day, she receives an email from her professor saying that her assignment failed to compile and that she needs to make changes and try again. Susan does not understand why it did not work on the professor's computer when it did on her own. The professor recommends that she install \pkg{fertile} and run \func{proj\_check} on her assignment. She does this and gets a message informing her that she used an absolute path to open her dataset when she should have use a relative path instead. She looks up what this means and then uses the new information to update her assignment. Her second submission compiles successfuly.

On future projects, she always loads and runs \pkg{fertile} before submitting.

\subsubsection{Experienced \R User}

Emma is a post-doc with several years of \R experience. She is familiar with some basic rules of reproducibility---file paths should always be relative and randomness should always be associated with a seed---but has never needed to pass any sort of reproducibility check before because her professors never emphasized that.

She has just finished a research project and is looking to submit her work to a journal. When researching the journal to which she is interested in submitting, she discovers that it has high standards for research reproducibility and a dedicated editor focusing on that aspect of submission. She goes online and finds the journal's guidelines for reproducibility. They are more complete than any guidelines to which she has previously been required to conform. In addition to notes about file paths and randomness, the journal requires a clean, well-organized folder structure, broken down by file category and stripped of files that do not serve a purpose. In order to be approved, submissions must also have a clear build chain and an informative \cmd{README} file.

Unsure of the best way to achieve this structure, Emma goes online to find help. In her search, she comes across \pkg{fertile}. She downloads the package, and in only a handful of commands, she identifies and removes excess files in her directory and automatically organizes her files into a structure reminiscent of an \R package. She now meets the guidelines for the journal and can submit her research.

\section{Results}

\pkg{fertile} is designed to: 1) be simple enough that users with minimal \R experience can use the package without issue, 2) increase the reproducibility of work produced by its users, and 3) educate its users on why their work is or is not reproducible and provide guidance on how to address any problems.

To test \pkg{fertile}'s effectiveness, we began an initial randomized control trial of the package on an introductory undergraduate data science course at Smith College in Spring 2020 \footnote{This study was approved by Smith College IRB, Protocol \#19-032}.

The experiment was structured as follows:

\begin{itemize}[noitemsep]

\item Students are given a form at the start of the semester asking whether they consent to participate in a study on data science education. In order to successfully consent, they must provide their system username, collected through the command \cmd{Sys.getenv("LOGNAME")}. To maintain privacy the results are then transformed into a hexadecimal string via the \func{md5} hashing function. 

\item These hexadecimal strings are then randomly assigned into equally sized groups, one experimental group that receives the features of \pkg{fertile} and one group that receives a control.

\item The students are then asked to download a package called \pkg{sds192} (the course number and prefix), which was created for the purpose of this trial. It leverages an \func{.onAttach} function to scan the \R environment and collect the username of the user who is loading the package and run it through the same hashing algorithm as used previously. It then identifies whether that user belongs to the experimental or the control group. Depending on the group they are in, they receive a different version of the package.

\item The experimental group receives the basic \pkg{sds192} package, which consists of some data sets and \R Markdown templates necessary for completing homework assignments and projects in the class, but also has \pkg{fertile} installed and loaded silently in the background. The package's proactive features are enabled, and therefore users will receive warning messages when they use absolute or non-portable paths or attempt to change their working directory. The control group receives only the basic \pkg{sds192} package, including its data sets and \R Markdown templates. All students from both groups then use their version of the package throughout the semester on a variety of projects.

\item Both groups are given a short quiz on different components of reproducibility that are intended to be taught by \pkg{fertile} at both the beginning and end of the semester. Their scores are then compared to see whether one group learned more than the other group or whether their scores were essentially equivalent. Additionally, for every homework assignment submitted, the professor takes note of whether or not the project compiles successfully.

\end{itemize}

Based on the results, we hope to determine whether \pkg{fertile} was successful at achieving its intended goals. A lack of notable difference between the \emph{experimental} and \emph{control} groups in terms of the number of code-related questions asked throughout the semester would indicate that \pkg{fertile} achieved its goal of simplicity. A higher average for the \emph{experimental} group in terms of the number of homework assignments that compiled successfully would indicate that \pkg{fertile} was successful in increasing reproducibility. A greater increase over the semester in the reproducibility quiz scores for students in the \emph{experimental} group compared with the \emph{control} group would indicate that \pkg{fertile} achieved its goal of educating users on reproducibility. Success according to these metrics would provide evidence showing \pkg{fertile}'s benefit as tool to help educators introduce reproducibility concepts in the classroom.

Unfortunately, we were unable to complete the analysis as intended as the trial had to be postponed after the COVID-19 pandemic significantly altered the experimental conditions at the midpoint of testing. Although the experiment was unsuccessful in its first attempt, we hope to run the same trial again and gather data on \pkg{fertile}'s effectiveness.

\section{Conclusion}\label{sec:conclusion}

\pkg{fertile} is an \R package that lowers barriers to reproducible data analysis projects in \R, providing a wide array of checks and suggestions addressing many different aspects of project reproducibility, including file organization, file path usage, documentation, and dependencies.
\pkg{fertile} is meant to be educational, providing informative error messages that indicate why users' mistakes are problematic and sharing recommendations on how to fix them. The package is designed in this way so as to promote a greater understanding of reproducibility concepts in its users, with the goal of increasing the overall awareness and understanding of reproducibility in the \R community.

The package has very low barriers to entry, making it accessible to users with various levels of background knowledge. Unlike many other \R packages focused on reproducibility that are currently available, the features of \pkg{fertile} can be accessed almost effortlessly. Many of the retroactive features can be accessed in only two lines of code requiring minimal arguments and some of the proactive features can be accessed with no additional effort beyond loading the package. This, in combination with the fact that \pkg{fertile} does not focus on one specific area of reproducibility, instead covering (albeit in less detail) a wide variety of topics, means that \pkg{fertile} makes it easy for data analysts of all skill levels to quickly gain a better understanding of the reproducibility of the work.

In the moment, it often feels easiest to take a shortcut---to use an absolute path or change a working directory. However, when considering the long term path of a project, spending the extra time to improve reproducibility is worthwhile. \pkg{fertile}'s user-friendly features can help data analysts avoid these harmful shortcuts with minimal effort.

\subsection{Future Work}\label{sec:future}

\pkg{fertile}, in its current version, addresses the vast majority of the aspects of reproducibility identified in Section \ref{sec:papers} in some way. However, there are several areas where further development to extend the available features of the package would be beneficial. These include the following:

\begin{itemize} [noitemsep]
  \item Expanding dependency management features to include \R session information and package version numbers in addition to package names.
  \item Expanding code and documentation style features to analyze whether code has been properly commented in addition to checking for a \cmd{README} and tidy code style.
  \item Adding \cmd{make}-like functionality that can analyze an \R project structure and files and use this information to generate a Makefile. This Makefile would have information about target files and their prerequisites and would assist with making sure that re-running an analysis is done as quickly as possible by ensuring that only the necessary code and files that have been updated are run when rebuilding and re-running code.
\end{itemize}

\section*{Data Availability Statement}
The sample project \cmd{project\_miceps} and package code associated with this paper can be found in the \cmd{R} and \cmd{tests} folders at \url{https://github.com/baumer-lab/fertile}.

\bibliographystyle{agsm}
\bibliography{refs}%

\end{document}